\documentclass[prb,twocolumn,superscriptaddress,showpacs,floatfix]{revtex4-1}

\usepackage{graphicx}
\usepackage{dcolumn}
\usepackage{amsmath}
\usepackage{color}

\begin{document}
\title{Alternating spin-orbital order in tetragonal Sr$_2$VO$_4$}

\author{M.~V.~Eremin}
\affiliation{Institute for Physics, Kazan (Volga region) Federal
University, 430008 Kazan, Russia}
\author{J.~Deisenhofer}
\affiliation{Experimentalphysik V, Center for Electronic
Correlations and Magnetism, Institute for Physics, Augsburg
University, D-86135 Augsburg, Germany}

\author{R.~M.~Eremina}
\affiliation{E. K. Zavoisky Physical Technical Institute, 420029
Kazan, Russia}


%
\author{J.~Teyssier}
\affiliation{D\'{e}partement de Physique de la Mati\`{e}re
Condense안, Universit\'{e} de Gen\`{e}ve, CH-1211 Gen\`{e}ve 4,
Switzerland}
\author{D.~van der Marel}
\affiliation{D\'{e}partement de Physique de la Mati\`{e}re
Condense안, Universit\'{e} de Gen\`{e}ve, CH-1211 Gen\`{e}ve 4,
Switzerland}
\author{A.~Loidl}
\affiliation{Experimentalphysik V, Center for Electronic
Correlations and Magnetism, Institute for Physics, Augsburg
University, D-86135 Augsburg, Germany}

\date{\today}

\begin{abstract}
Considering spin-orbit coupling, the tetragonal crystal-field, and
all relevant superexchange processes including quantum interference,
we derive expressions for the energy levels of the vanadium ions in
tetragonal Sr$_2$VO$_4$. The used parameters of the model
Hamiltonian allow to describe well the excitation spectra observed
in neutron scattering and optical experiments at low temperatures.
The free energy exhibits a minimum which corresponds to a novel
alternating spin-orbital order with strong thermal fluctuation of
the orbital mixing parameter.

\end{abstract}



\maketitle

In many transition-metal compounds the orbital degrees of freedom
play a decisive role in determining the ground-state properties of
materials such as manganites, titanates, or
vanadates.\cite{Tokura00} When contributions of the orbital moment
and spin-orbit coupling are not negligible, a separation between
spin and orbital degrees of freedom is not adequate anymore and the
system is better described by an effective total angular
momentum.\cite{Abragam70,Kant08} If spin-orbit coupling competes
with electron-phonon or exchange interactions even strong
fluctuation regimes can arise.\cite{Feiner82,Tsurkan10,Wang11}

The system investigated here is the layered insulator Sr$_2$VO$_4$
with tetragonal symmetry, which early on has come into focus as an
isostructural $d^1$ analogue of La$_2$CuO$_4$.\cite{Pickett89}
Consequently, it was suggested that Sr$_2$VO$_4$ could become
superconducting upon applying chemical pressure by doping or
external pressure.\cite{Singh91,Arita07} While the system could not
be driven towards superconductivity, it turned out to be a model
system for studying the interplay of orbital-lattice, spin-orbital
and superexchange interactions.\cite{Imai05,Zhou07, Jackeli09,
Zhou10}

\begin{figure}[t]
\includegraphics[width=0.45\textwidth,clip]{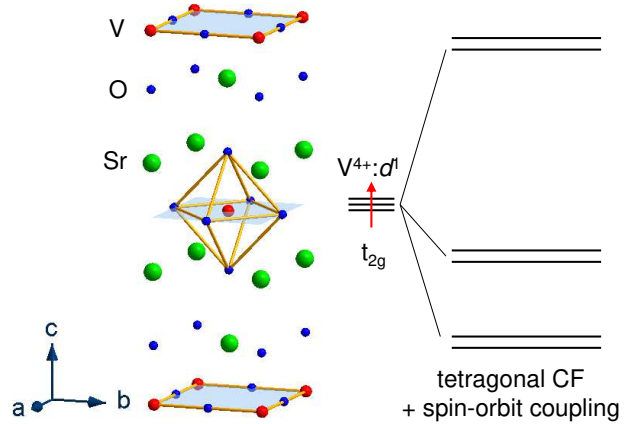}
\caption{\label{fig:structure} Left: Unit cell of tetragonal Sr$_2$VO$_4$ with symmetry I4/mmm (Ref.~\onlinecite{Rey90})
highlighting the VO$_2$ planes and the octahedral coordination of the V ions.
Right: Splitting of the V$^{4+}$ $t_{2g}$-levels due the tetragonal crystal field and spin-orbit coupling.}
\end{figure}

In tetragonal Sr$_2$VO$_4$  with space group
$I_4/mmm$,\cite{Cyrot90,Rey90} the octahedrally coordinated V$^{4+}$
ions occupy a square lattice in the $ab$-plane (see
Fig.~\ref{fig:structure}). The magnetic ground state has been
claimed to be antiferromagnetic with transition temperatures in the
range 10 - 100~K determined from susceptibility measurements, but
long-range order remained evasive on the basis of
neutron-diffraction studies.\cite{Cyrot90,Nozaki91,Suzuki92} Recent
studies established the occurrence of a magneto-structural phase
transition extending over a temperature range from 94 - 122~K. Both
the high-temperature and the low-temperature structure are
tetragonal and reportedly coexist within this range.\cite{Zhou07}
Specific heat data revealed two distinct broad maxima occurring at
98 and 127~K mirroring the borders of the two-phase
regime.\cite{Viennois10,Teyssier11}

The disappearance of the high-temperature phase is accompanied by a
significant drop in the susceptibility at about 100~K, which has
been attributed to the onset of long-range AFM and orbital
order.\cite{Zhou07} Theoretically, the ground state of Sr$_2$VO$_4$
has been interpreted in terms of stripe-like orbital and collinear
AFM spin order,\cite{Imai05} or an ordering of magnetic
octupoles.\cite{Jackeli09} Inelastic neutron scattering revealed two
excitations at about 120~meV, which were assigned to the highest
lying doublet of the V$^{4+}$ $t_{2g}$ levels.\cite{Zhou10} Recent
optical experiments reported excitations at 31~meV (visible for $T
>$ 80~K), 36~meV (visible for $T <$ 120~K) and a two-peak structure
at 100~meV and 108~meV, which remains visible from 13~K to room
temperature.\cite{Teyssier11}

To elucidate the nature of the ground state and to understand the
observed excitation spectrum we consider the effects of spin-orbit
coupling, crystal-field, and superexchange on the energy levels of
the vanadium ions. The resulting free energy points towards a novel
alternating spin-orbital order in the ground state.


To describe the system of  V$^{4+}$ ions we use the Hamiltonian
$H=H_{si} + H_{ex}$ where $H_{ex}$ describes the exchange coupling
of neighbouring ions and $H_{si}$ contains the single-ion
contributions in a tetragonal crystal field:
\begin{equation}H_{si}=
D[3l_{z}^{2}-l(l+1)]+\lambda _{c}l_{z}s_{z}+\frac{\lambda _{a,b}}{2}%
(l_{-}s_{+}+l_{+}s_{-})
\end{equation}
Here $D$ denotes the single-ion anisotropy and $l$ the effective
angular momentum $l$ = 1 of the $t_{2g}$-orbitals, which we describe
using $|1\rangle=-\frac{1}{\sqrt{2}}\left[ d_{yz}+id_{xz}\right]$,
$|-1\rangle=\frac{1}{\sqrt{2}}\left[ d_{yz}-id_{xz}\right]$, and
$|0\rangle=d_{xy}$ as a basis.\cite{Abragam70} Moreover, we use
anisotropic spin-orbit coupling constants $\lambda_c$ and $\lambda
_{a,b}$ parallel and perpendicular to the $c$-direction. Anisotropic
spin-orbit coupling can arise due to covalency effects and has been
observed in several $d^1$ systems in octahedral
environment.\cite{Abragam70,Zakharov06}

The superexchange coupling between $V$ ions via oxygen ions in the
$ab$ plane is usually described via the corresponding hopping
integrals,\cite{Anderson} which in our case are given by:
\begin{eqnarray}
t_{1,1}&=&t_{-1,-1}=\frac{1}{2}\left( t_{xz,xz}+t_{yz,yz}\right)\\
t_{1,-1}&=&t_{-1,1}=\frac{1}{2}(t_{xz,xz}-t_{yz,yz})\\
t_{0,0}&=&t_{xy,xy}
\end{eqnarray}
From the spatial distributions of the $d_{xz}$ and $d_{yz}$ orbitals
it is clear that the signs of the transfer integrals $t_{xz,xz}$ and
$t_{yz,yz}$ are different and, therefore, $\left\vert
t_{1,-1}\right\vert >\left\vert t_{1,1}\right\vert$. This
observation will allow us to deduce the most likely ordering of the
V states in the ground state.

Using the reported crystal structure of Sr$_{2}$VO$_{4}$ one finds
that $D<0$ and, therefore, the possible ground states of the
V$^{4+}$ ions are $|\pm 1,\pm 1/2\rangle$. The antiferromagnetic
superexchange coupling (see below) will yield an additional gain in
energy when $|-1,\pm 1/2\rangle$- states are surrounded by $| 1,\mp
1/2\rangle,$ or vice versa. Then, keeping in mind that the
spin-orbit coupling parameters $\lambda _{c}<0$, we arrive at a
configuration in the $ab$ plane of Sr$_{2}$VO$_{4}$  where each
vanadium ion in the state $|1,1/2\rangle$  is surrounded by vanadium
ions in the $|-1,-1/2\rangle$ state and vice versa. According to the
third Hund rule, spin ($s_z=\pm 1/2$) and angular ($l_z=\mp 1$)
momentum of the V$^{4+}$ ground state configuration are in
opposition. The corresponding combined spin-angular moment per site
therefore possesses the peculiarity that the magnetic moment
$m_z/\mu_B = 2s_z - \kappa l_z$ is almost completely muted, when the
the covalency reduction factor $\kappa$ is close to
one.\cite{Abragam70} The resulting ordering scheme can be described
as an alternating order of spin and orbital moments on each site.

\begin{figure}[t]
\includegraphics[width=0.4\textwidth]{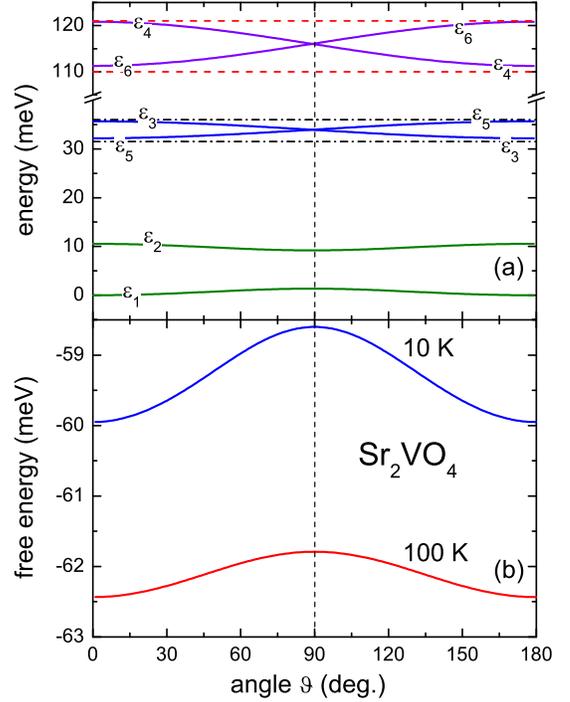}
\caption{\label{fig:levelscheme} (a) Calculated energy levels as a
function of the orbital mixing angle $\vartheta$ using Eqs.~(13)-(15) and
parameters $\lambda_c$=-30~meV, $\lambda_{ab}$=-28~meV, $D$=
-33~meV, $J_a$=-15~meV, $J_f$=-9.3~meV, $J_f^\prime$=-0.7~meV, and
$J_{int}$ = 7.5~meV. Excitation energies observed
by neutron scattering (Ref.~\onlinecite{Zhou10}) and optical spectroscopy(Ref.~\onlinecite{Teyssier11}) are shown as dashed and dash-dotted lines, respectively.
(b) Free energy as a function of $\vartheta$ calculated with the same parameters 10 and 100~K.}
\end{figure}


First, we introduce the superexchange parameters
\begin{eqnarray}
J_{a}&=&4\frac{t_{xz,xz}^{2}+t_{yz,yz}^{2}}{U}\\
J_{int}&=&-8\frac{t_{xz,xz}t_{yz,yz}}{U},
\end{eqnarray}
where $U$ denotes the onsite Coulomb repulsion. The signs of the
transfer integrals $t_{xz,xz}$ and $t_{yz,yz}$ are different and,
therefore, the cross term $J_{int}$\ is positive. This parameter
describes the quantum interference effect in superexchange coupling.
The part of the effective exchange Hamiltonian containing these
parameters is written as
\begin{eqnarray}
H_{ex}(1)&=&  \frac{J_{a}}{8}\left(
\mathbf{s}_{i}\mathbf{s}_{j}-\frac{1}{4}\right)
(2l_{iz}^{2}l_{jz}^{2}+l_{i+}^{2}l_{j+}^{2}+l_{i-}^{2}l_{j-}^{2})\nonumber\\
&-&\frac{ J_{int}}{4}\left(
\mathbf{s}_{i}\mathbf{s}_{j}-\frac{1}{4}\right)l_{iz}l_{jz}.
\end{eqnarray}

The ferromagnetic contributions to the superexchange interaction
comprise two exchange integrals $J_{f}$ and $J_{f}^{\prime}$, which
for a pair of V ions along the $x$-axis can be denoted as
\begin{eqnarray}
J_{f}&=&J_{p}-2\frac{I_f}{U^{2}}\left( t_{xy,xy}^{2}+t_{xz,xz}^{2}\right) \quad \mbox{and}  \\
J_{f}^{\prime}&=&J_{p}^\prime-2\frac{I_{f}}{U^{2}}\left( t_{xz,xz}^{2}\right)\label{jfprime} \quad \mbox{with}\\
I_f&=&\langle d_{xy},d_{xz}|\frac{e^2}{r_{12}}|d_{xz},d_{xy}\rangle. \nonumber
\end{eqnarray}

Here $I_f$ is an exchange integral, which can be estimated via Racah
parameters as $3B + C$~=~0.9~eV and $J_{p},J_{p}^\prime$ correspond
to potential exchange contributions. In Eq.~(\ref{jfprime}) the
potential exchange parameter $J_{p}^\prime$ is expected to be small
and will be neglected in the following. Note that $J_{f}^{\prime }$
has been called Hunds-coupling parameter in
Ref.~\onlinecite{Jackeli09}.

In momentum representation for $l=1,$ $s=1/2$ this part of the
exchange Hamiltonian is written as

\begin{widetext}
\begin{eqnarray}
H_{ex}(2_{x})=\left(
\mathbf{s}_{i}\mathbf{s}_{j}+\frac{3}{4}\right) \left[J_{f}\left(
-2+l_{ix}^{2}+l_{jx}^{2}+l_{iz}^{2}l_{jy}^{2}+l_{iy}^{2}l_{jz}^{2}\right)
+J_{f}^{\prime }(l_{ix}^{2}+l_{jx}^{2}-2l_{ix}^{2}l_{jx}^{2})\right]
\end{eqnarray}
\end{widetext}
The effective Hamiltonian for a pair along the $y$-axis can be
obtained by a permutation of indices $x\longrightarrow y,$
$y\longrightarrow x$. For a detailed discussion of the differences
in spin dependent factors of ferro- and antiferromagnetic exchange
terms we refer to Ref.~\onlinecite{Eremin81}. Let us consider now a
two-sublattice configuration in which each V ion of sublattice $i$
is described by the wave function $|\vartheta\rangle=\cos
\vartheta/2|1,1/2\rangle+\sin \vartheta/2 |-1,-1/2\rangle$ with
$\vartheta= \vartheta _{i}$ is surrounded by four V ions of
sublattice $j$ with $\vartheta= \vartheta _{j}$.\cite{footnote}
Using the effective exchange operator $H_{ex}=\sum
[H_{ex}(1)+H_{ex}(2_{x})+H_{ex}(2_{y})]$ and assuming that only the
ground states of surrounding V ions are populated we arrive at the
following energy spectrum of the vanadium ions:

\begin{widetext}


\begin{equation}
\varepsilon _{1,2}=D+\frac{\lambda _{c}}{2}-\frac{J_{a}+J_{int}}{4}
+\frac{3J_{f}^{^{\prime }}}{2}\pm \frac{1}{4}\left[
(J_{a}+J_{int}+2J_{f}^{^{\prime }})u^{2}+(J_{a}-J_{f}^{^{\prime }})^{2}v^{2}
\right] ^{\frac{1}{2}} \label{eq:levels12}
\end{equation}

\begin{eqnarray}
\varepsilon _{3,4}&=&-\frac{D}{2}-\frac{\lambda _{c}}{4}+\frac{J_{a}-J_{int}}{8}(u-1)+\frac{J_{f}}{4}(3-u)+\frac{3J_{f}^{^{\prime
}}}{2}\nonumber\\
&&\pm \frac{1}{2}\left[ \left(3D-\frac{\lambda _{c}}{2}+\frac{J_{a}-J_{int}}{4}
(u-1)-\frac{J_{f}}{2}(3-u)+J_{f}^{^{\prime }}u\right)^{2}+2\lambda _{a,b}^{2}%
\right]^{\frac{1}{2}} \label{eq:levels34}
\end{eqnarray}

\begin{eqnarray}
\varepsilon _{5,6}&=&-\frac{D}{2}-\frac{\lambda _{c}}{4}-\frac{
J_{a}-J_{int}}{8}(u+1)+\frac{J_{f}}{4}(3+u)+\frac{3J_{f}^{^{\prime
}}}{2}\nonumber\\
&&\pm \frac{1}{2}\left[ \left(3D-\frac{\lambda _{c}}{2}-\frac{J_{a}-J_{int}}{4}
(u+1)-\frac{J_{f}}{2}(3+u)-J_{f}^{^{\prime }}u\right)^{2}+2\lambda _{a,b}^{2}
\right] ^{\frac{1}{2}} \label{eq:levels56}
\end{eqnarray}


\end{widetext}

Here we have introduced $u=\cos \vartheta _{j}$ and $v=\sin
\vartheta _{j}$. From the expressions for $\varepsilon _{3,4}$ and
$\varepsilon _{5,6}$ one finds that at $\vartheta =\pm \pi /2$ the
excited states are degenerate doublets, i.e. $\varepsilon
_{3}=\varepsilon _{5}$ and $\varepsilon _{4}=\varepsilon _{6}$.
However, this choice of $\vartheta =\pm \pi /2$ cannot explain the
observed splitting of the highest-lying doublet $\varepsilon _{4,6}$
which was observed by neutron scattering and optical
spectroscopy.\cite{Zhou10,Teyssier11}

Now let us turn to the ground state $\varepsilon _{1}$. A minimum in
energy of this level will occur at $\vartheta =\pm \pi /2$ only if
$3\vert J_{f}^{^{\prime }}\vert >J_{int}$. If $J_{int}>3\vert
J_{f}^{^{\prime }}\vert$ the minimum will occur at $\vartheta =0$ or
$\vartheta =\pi $. At these angles the excited states are split due
to the exchange-molecular field in agreement with
experiment.\cite{Zhou10} Using the experimentally observed
splitting\cite{Zhou10} of the highest doublet of about 10~meV we
estimate the value $\vert J_{f}+J_{f}^{\prime }\vert \simeq$ 10~meV.
Following Imai and coworkers the energy cost for moving a
3d-electron between V ions in Sr$_{2}$VO$_{4}$ is about $U\simeq$
11~eV and the effective transfer integrals $t_{xz,xz}\simeq$-0.2~eV
and $t_{yz,yz}\simeq$ 0.05~meV.\cite{Imai05} Therefore, we estimate
$J_{a}\simeq$ 15~meV, $J_{int}\simeq$ 7.5~meV, $ J_{f}^{\prime
}=-(3B+C)/2U*J_{a}\simeq -0.7$~meV, and $ J_{f}\simeq -9.3$~meV. The
values $D$ = -33~meV, $\lambda_{c}$ = -30~meV, and $\lambda_{ab}$ =
-28~meV are in agreement with conventional
estimates.\cite{Abragam70,Jackeli09} Using these values we plot the
energy levels as
$\varepsilon_i(\vartheta)-\varepsilon_1(\vartheta=0,\pi)$ in
Fig.~\ref{fig:levelscheme}(a) as a function of the orbital-mixing
angle $\vartheta$. Note that $\varepsilon_1(\vartheta)$ is not
constant but becomes minimal for $\vartheta=0,\pi$. The estimated
excitation energies $\varepsilon_4-\varepsilon_1$ = 121~meV,
$\varepsilon_6-\varepsilon_1$ = 111~meV, and
$\varepsilon_3-\varepsilon_1$= 36~meV for $\vartheta$ = 0 (or
corresponding values for $\vartheta = \pi$)  are in agreement with
the transitions observed by optical (dash-dotted lines) and neutron
scattering experiments (dashed lines) at low temperatures shown in
Fig.~\ref{fig:levelscheme}(a).\cite{Zhou10,Teyssier11}

\begin{figure}[t]
\includegraphics[width=0.40\textwidth,clip]{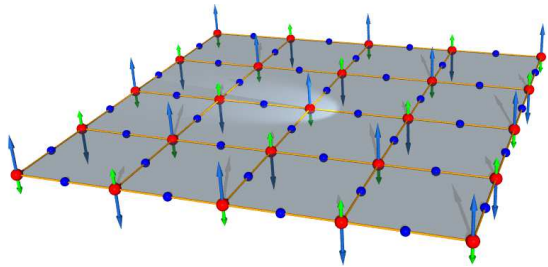}
\caption{\label{fig:plane} Sketch of the proposed alternating spin-orbital order in
the $ab$-plane of tetragonal Sr$_2$VO$_4$. Short and long arrows correspond to spin and orbital moments of the V ions, respectively.}
\end{figure}

In Fig.~\ref{fig:levelscheme}(b) we compare the free energy per
vanadium site at 10 and 100~K (close to $T_N$) as a function of the
parameter $\vartheta$ using the values for the exchange constants
estimated above:
\begin{equation}
F(T,\vartheta )=-k_{B}T\ln\sum \exp \left( -\frac{\varepsilon
_{i}(\vartheta )}{k_{B}T}\right)
\end{equation}
There is a minimum at $\vartheta =0$ and $\vartheta =\pm \pi$ in
both cases and the energy difference with respect to $\vartheta =\pm
\pi /2$ decreases from 1.5~meV to 0.5~meV, respectively. Hence, one
can anticipate that with increasing temperature considerable
fluctuations of $\vartheta$ can be expected. We estimate these
fluctuations by $\overline{\Delta \vartheta^2}$ = $k_BT(\partial^2
F/\partial\vartheta^2)^{-1}\simeq (\pi/5)^2$ and $(\pi/3)^2$ for 10
and 100~K respectively. Certainly, the proposed spin-orbital ordered
state as depicted in Fig.~\ref{fig:plane} will be destabilized at
high temperatures. However, the splitting of the highest-lying
doublet remains almost unchanged across the Neel
temperature,\cite{Zhou10} and it is reasonable to expect the
exchange splitting of $\varepsilon_{3,5}$ to survive as well, even
though the exchange parameters might be somewhat reduced. We would
like to mention that the value $\varepsilon_5-\varepsilon_1$= 31~meV
corresponds nicely to the optical excitation observed for $T>$ 80~K,
but given the strong fluctuations expected for this temperature
range the interval
$(\varepsilon_3+\varepsilon_5)/2-(\varepsilon_2+\varepsilon_1)/2
\sim$ 29~meV might provide a more suitable estimation of the
high-temperature optical excitation.


In summary, we calculated the level-scheme for the energy levels of
the vanadium ions and propose an alternating spin-orbital ordering
with almost muted magnetic moment as the ground state for
Sr$_2$VO$_4$. The proposed scenario and parameter values allow to
obtain a consistent picture of the low-temperature excitation
spectrum of Sr$_2$VO$_4$, which was recently reported by neutron and
optical experiments.

We thank G.~Jackeli, D.I.~Khomskii, and H.-A.~Krug von Nidda for
stimulating discussions. This work is partially supported by the
SNSF and the National Center of Competence in Research (NCCR) MaNEP
and by the DFG via the Collaborative Research Center TRR 80
(Augsburg-Munich). MVE acknowledges support by Ministry Education of
the Russian Federation via Grant No. 1.83.11.



\begin{thebibliography}{39}
\bibitem{footnote} Note that the orbital mixing angle $\vartheta$ used in our approach is
different from the notation introduced in
Ref.\onlinecite{Jackeli09}.
\end{thebibliography}


\end{document}